
\magnification 1200
\overfullrule=0pt
\def\etal{{\sl et al.}}

\centerline{\bf Gravitational Waves and $\gamma$-Ray Bursts}
\medskip
\centerline{ \it Christopher S. Kochanek\footnote{$^{[a]}$}{Alfred P. Sloan
  Foundation Fellow} }
\medskip
\centerline{ \it and }
\medskip
\centerline{ \it Tsvi Piran\footnote{$^{[b]}$}{Permanent Address: Racah
  Institute for Physics, Hebrew University, Jerusalem, Isreal} }
\medskip
\centerline{ Harvard-Smithsonian Center for Astrophysics }
\centerline{ 60 Garden Street }
\centerline{ Cambridge, MA 02138, USA }

\bigskip

{ \bf Coalescing binaries in distant galaxies are one of the most
promising sources of gravitational waves detectable by the LIGO
project.$^{[1-5]}$ They are also a copious source of
neutrinos,$^{[1]}$ however these neutrino pulses are far too weak to
be detected on earth. Several years ago Eichler \etal$\,$$^{[6]}$
suggested that they are also sources of $\gamma$-ray bursts (GRBs).
Recently it was found$^{[7]}$ that GRBs are likely to be cosmological
in origin, and coalescing binary systems$^{[6,8-11]}$ are probably the
most promising cosmological sources. The current estimates of the
burst and LIGO rates from a cosmologically distributed population are
based on the systems observed in our galaxy.$^{[12-14]}$ These
estimates are based on only three binarys so there are large
statistical uncertainties in the coalescence rate.  If we accept the
cosmological/coalescing binary hypothesis for GRBs, then we get a more
accurate estimate of the rate at which binaries coalesce and hence of
the predicted LIGO signal.  The association between GRBs and binaries
can significantly improve the performance of LIGO.  The detection of
gravitational radiation from a GRB source not only confirms the
coalescing binary model, but it will also provide information on the
geometry and energy generation mechanism of the burst.}

GRO$^{[7]}$ has shown that GRBs are located in the outer parts of
the galactic halo or cosmologically. In both cases all GRB
mechanisms must evolve through a pair fire ball phase,$^{[15]}$
which makes it difficult to distinguish
between types of sources based on burst characteristics.
The association between GRBs and LIGO sources is an unambiguous,
verifiable prediction of the coalescing binary model,$^{[16]}$
unlike gravitational lensing$^{[17]}$ and redshift effects
which are generic to all cosmological models.

The ability of LIGO to find coalescing binaries is strongly limited by
the random noise levels in the detectors and the need to continuously search
for weak signals.  There is a significant gain in sensitivity if the
search for signals is limited to narrow windows of time near bright
$\gamma$-ray bursts.
The polarization of the gravitational waves provides evidence about the
emission geometry of the GRB sources, because the relative strengths of
the $+$ and $\times$ polarizations is a function of the inclination angle.
If the $\gamma$-rays are preferentially emitted along the polar axis of
the binary, then the GRB sources are stronger LIGO sources than the average
coalescing binary because binaries emit gravitational waves more strongly
along the polar axis than in the orbital plane.

The evidence for a cosmological origin of GRBs is the combination of
isotropy on the sky and inhomogeneity in space.  The $C/C_{min}$
distribution for bright bursts has a $-3/2$ slope characteristic of
a homogeneous, Euclidean distribution.  Fainter bursts show a sharp break from
this slope, indicating that the distribution becomes inhomogeneous at large
distances.   The shape of the $C/C_{min}$ for
cosmological sources is determined by redshift effects that change the
luminosity, and the effective volume.$^{[16,18-20]}$
The relative sharpness of the break suggests the GRBs are
good standard candles, because a wide intrinsic luminosity distribution
would have smeared out the break.$^{[18]}$

We consider a burst with a differential luminosity per unit energy $E$
equal to $ (dL/dE) = L_0 E^{-\alpha}$. In a detector like BATSE the
number of counts, $C$, during a given  time interval, $\Delta t$,
from redshift $z=x-1$ is
$$ C =
 {(  L_0 /\alpha) E_{min}^{-\alpha} x^{(1-\alpha)}  \over   4 \pi D_L^2(z)},
         \eqno(1)
$$
where $D_L$ is the luminosity distance and $E_{min}$ is the minimum detectable
energy.
If the comoving rate of coalescences is $r(z)$ per unit comoving volume, then
the integrated rate of coalescences closer than redshift $z=x-1$ is
$$
R(z) = 4 \pi r_H^3 \int_1^{1+z} x^{-9/2} \hat{D}_L^2 r(z) dx \eqno(2)
$$
where $D_L = r_H \hat{D}_L=2 r_H (x-x^{1/2})$ for the Einstein-DeSitter
cosmology we use in this study, $r_H=c/H_0$  is the Hubble radius, and
$H_0 = 100 h_0 $ km s$^{-1}$ Mpc$^{-1}$.

For a given spectral index $\alpha$ and evolution model $r(z)$, the
observed distribution of $C/C_{min}$ is fit to infer the
maximal redshift, $z_{max}$, from which GRO detects GRBs.$^{[16]}$
For a constant coalescence rate, $r$, and spectral index
$\alpha=1$, the best fit to the $C/C_{min}$ distribution gives
$z_{max}=1.45$.  Varying the spectral index from 0 to 2 changes
$z_{max}$ from 2.8 to 1.  The dominant uncertainty lies in the
comoving coalescence rate, $r(z)$, since a very strong evolution
of the event rate $r(z) \propto t(z)^2$ with cosmic time $t(z)$
can reduce $z_{max}$ to as low as $z_{max}=0.5$.

The observed GRB rate is $\approx 800$ events per year$^{[7]}$
which corresponds to a local merger rate of
$3.6 \times 10^{-8} h_0^{-3}$ Mpc$^{-3}$ yr$^{-1}$ or
$3.6 \times 10^{-6} n_2^{-1} $ mergers per galaxy per year
for a galaxy density of $n=10^{-2} h_0^{-3} n_2$ Mpc$^{-3}$.
The current rate increases by a factor of 10 if
we assume a strong cosmological time dependence ($r(z) \propto
t^2(z)$).  The overall agreement between estimated merger rate (based
on GRBs) and the estimates of
$10^{-5\pm 1}$ yr$^{-1}$ mergers in the Galaxy
based on galactic binary pulsars$^{[13,14]}$
supports the coalescing neutron star model as sources for GRBs.$^{[16]}$
The agreement is better if we assume that the event rate is
independent of cosmological time. This suggests that we use the
calculated distribution of GRBs to predict the distribution of
gravitational wave strains $h_c$, that could be observed by LIGO.

The characteristic strain produced by a coalescing binary with reduced
mass $\mu = 1.4 \mu_{NS} M_\odot$ and total mass $M = 2.8 M_{NS} M_\odot$
scaled to the masses characteristic of NS-NS binaries,
averaged over binary inclinations, and using an optimal filter$^{[3,21]}$
with a knee at frequency $f=100 f_{c2} $ Hz to detect the signal is
$$ h_c = 2.3 \times 10^{-23} h_0 \mu_{NS}^{1/2} M_{NS}^{1/3}
      f_{c2}^{-1/6} { r_H x^{5/6} \over D_L }.
      \eqno(3)
$$
This is not the same  as the instantaneous strain at frequency $f$,
which varies as $r_H x^{5/3} / D_L$, because the optimal filter
effectively integrates the signal over many orbital periods to
increase the signal to noise ratio.

In Figure 1 we show the expected rate of events stronger than a given
strain $h_c$ for three different models: a constant comoving rate of
$r=10^{-7}h^{-3}$ Mpc$^{-3}$ yr$^{-1}$ based on the galactic
estimates,$^{[13,14]}$ the rate
using the fit to the $C/C_{min}$ distribution with $\alpha=1$
and a constant comoving rate for which $z_{max}=1.45$,
and the rate using the fit to the $C/C_{min}$
distribution with $\alpha=1$ and $r(z) \propto t(z)^2$
for which $z_{max}=0.5$.  The rapidly evolving model may be inconsistent
with the local galactic estimates.

The signal processing for the LIGO project is predicated on using two or more
detectors to eliminate non-Gaussian sources of noise.  Under this assumption,
the instantaneous noise level in the LIGO detector$^{[3,21]}$ is given by
$h_n^2 = f_c S_n(f_c)/ \langle F_+^2 \rangle$ where $f_c$ is the knee
frequency, $\langle F_+^2 \rangle =1/5$ is the angle and polarization averaged
detector beam pattern, and $S_n(f_c)$ is the spectral density of the noise
at the knee frequency $f_c$.
This is only the average instantaneous noise level, and to estimate the
detectability of a source we must take into account the fluctuations of the
noise.  If we are using optimal filtering, we must make trial
correlations of the filter with the detector output approximately
once every $f_c^{-1}=0.1 f_{c2}^{-1}$ seconds, or about
$3 \times 10^8 f_{c2}^{-1}$ trials per year.  For two such Gaussian
detectors in which we expect 3 events per year, we must have a
signal to noise ratio in the signal of $S^2/N^2 = \ln (10^8 f_{c2}^{-1})$
to be 90\% confident the detectors are not seeing noise.  This defines
the detection threshold for events with an expected rate of  3 per
year$^{[3]}$,
$h_{3/yr} = (\ln[ 10^8 f_{c2}^{-1} ])^{1/2} h_n (f_c) \simeq 4.2 h_n(f_c)$.
This estimate is insensitive to changes to 99\% confidence rather
than 90\%, or 0.3 events per year rather than 3 because of the logarithmic
dependence. Using the estimates from $[3]$,
the initial version of LIGO should have a noise level of
$h_{3/yr} \simeq 8 \times 10^{-20} f_{c3}^{3/2}$ for detecting bursts with
knee frequencies near $f=1 f_{c3} $ kHz.  The higher knee frequency
also reduces the strength of the signal by a factor of $1.5$ because
of the $f^{-1/6}$ dependence in $h_c$.  The advanced versions of
LIGO should have better low frequency sensitivities and noise levels
of $h_{3/yr} \simeq 1.3 \times 10^{-22} f_{c2}$ for detecting bursts with
knee frequencies near $f=100 f_{c2}$ Hz.  The quantum limit for detecting
bursts is $h_{3/yr} \simeq 1.6 \times 10^{-25} f_{c2}^{-1/2} $.

This is not true if we use $\gamma$-ray bursts to trigger searches for
gravitational waves.  As we can see in Figure 1, LIGO is sensitive only
to signals from the rare, bright bursts, so we should search
only a short time interval near the bright bursts for gravitational waves.
We expect the GRB to occur between several dynamical times (milliseconds)
and several cooling times (seconds) after the coalescence of the binaries.
If we are conservative and search for one minute before each of the ten
brightest bursts in a given year, then we are
conducting only $6000 f_{c2}^{-1}$ trials rather than $10^8 f_{c2}^{-1}$
trials, and the detection threshold for $\gamma$-ray burst events is
$1.5$ times larger than the $h_{3/yr}$ estimates.  An increase in the
detection threshold by a factor of $1.5$ increases the rate by a factor
of three.  Events correlated with GRBs not only come at a known time,
but also from a known direction.  This reduces the number of trial correlations
further, but it is not as important as restricting the search in
time because of the poor spatial resolution of LIGO.

If the use of two detectors does not eliminate all non-Gaussian errors in
the LIGO signal processing,
then the use of Gaussian statistics to estimate $h_{3/yr}$ will fail
catastrophically for the random search mode, because the noise estimate
is working way out on the tail of the Gaussian distribution.  If the
noise has a non-Gaussian tail, triggered searches using the GRBs may
be the only way LIGO can detect coalescing binaries.

Coalescing binaries emit gravitational waves much more strongly along
the polar axis than in the orbital plane, with $h_+ \propto (1+\cos^2
i)$ and $h_\times \propto 2 \cos i$ where $i$ is the inclination angle
of the binary.$^{[22]}$  The characteristic amplitude
$h_c \propto \langle |h_+|^2+|h_\times|^2\rangle^{1/2} $ includes an
average over the inclination angle between the observer and the
binary.  If the $\gamma$-ray emission from a coalescing binary is
beamed along the polar axis, we have underestimated the average
gravitational wave signal by the factor
$(1+11x/16+11x^2/16+x^3/16+x^4/16)^{1/2}$ where $x=\cos\psi$ and
$\psi$ is the opening angle around the normal into which the bursts
are beamed. If the GRB is associated with a disk formed during the
merger$^{[10,11]}$ then we would expect
the emission to be beamed along the polar axis.  The maximum gain is a
factor of $1.6$ in $h_c$, and the gain is $1.2$ for $\psi=30^\circ$.
Because the detection volume increases as the cube of the detection
limit on the strain, the rate increases are factors of 4 and 1.9
respectively over unbeamed bursts.  If the bursts are detected by LIGO
the beaming can be detected from the polarization of the bursts:
unbeamed bursts tend to be dominated by the $+$ polarization, while
beamed bursts are dominated by the $\times$ polarization of the
gravitational waves.  If the bursts are beamed along the polar axis,
then the GRB rate misses fraction $\cos\psi$ of the coalescences.

The hypothesis associating GRBs with coalescing binaries has several
important implications for gravitational wave searches using LIGO.
The GRB rate can accurately estimate the rate of coalescences as a
function of the gravitational wave strain on the earth.  The LIGO
sensitivity must have a strain sensitivity $h_c$ greater than
$10^{-21.1} h_0$ for an event rate of one per decade, and greater than
$10^{-21.4} h_0$ for a rate of once per year.  The strain sensitivity
of LIGO is considerably greater if the brightest GRBs serve as
triggers to search for gravitational waves in a narrow window of time
near the burst, and in the direction of the burst.  The strain
sensitivity is roughly 50\% greater in a coincidence experiment
compared to a random search, and the detection rate is roughly three
times higher.

The presence or absence of coincidences between GRB and LIGO events
may be the final proof or disproof of the coalescing binary
hypothesis.  Once the strain sensitivity reaches $10^{-20.7} h_0$ LIGO
can begin to set limits on GRB models, initially by setting limits on
the redshift distribution.  If the sensitivity reaches $10^{-21.5}
h_0$ then either LIGO begins to detect coalescing binaries in
coincidence with GRBs, or the coalescing binary model is dead.  If
they are related, then polarization of the gravitational waves can be
used to determine the emission geometry of the GRBs.  The time delay
between the merging of the binary and the GRB will help to determine
the emission mechanism.  The shortest possible delays will be
comparable to the dynamical time (milliseconds), and longer delays
might be viscous or weak interaction time scales.

The burst rate is an underestimate of the LIGO rate.  For a given
class of binaries, beaming of the $\gamma$-rays may mean that only a
fraction of the coalescences are seens as GRBs.  Moreover, not all
types of coalescing compact binaries may cause bursts.  The galactic
rates are only estimated from NS-NS binaries, and there may be a
comparable number of NS-BH mergers and a much smaller number of BH-BH
mergers.$^{[13]}$ It is not clear whether NS-NS or NS-BH
binaries are more likely to produce bursts, although the dynamical
stability of orbits and evolution time scales favor NS-NS binaries
over NS-BH binaries.$^{[23]}$

\medskip
\noindent{\bf Acknowledgements:} The authors thank Ramesh Narayan and
Shude Mao for helpful discussions.

\medskip
\def\ref{\par \smallskip \noindent \hangindent .5in \hangafter 1}
\def\ApJL{{\it Astrophys. J. Lett.}}

\noindent {\bf References}
\ref
[1] Clark, J.P.A., \& Eardley, D.M., 1977, {\it ApJ}, {\bf 212}, 311.
\ref
[2] Schutz, B.F., 1986, {\it Nature}, {\bf 323}, 310.
\ref
[3] Thorne, K.S., 1987, {\it 300 Years of Gravitation}, S.W. Hawking \&
  W. Isreal, eds., (Cambridge Univ. Press: Cambridge) 378.
\ref
[4] Abramovici, W.E. \etal  1992, {\it Science}, {\bf 256}, 325.
\ref
[5] Cutler, C., \etal, 1992, {\it Caltech preprint}.
\ref
[6] Eichler, D., Livio, M., Piran, T., and Schramm, D. N. 1989,
  {\it Nature}, {\bf 340}, 126.
\ref
[7] Meegan, C.A., \etal, {\it Nature}, {\bf 355}, 143.
\ref
[8] Piran, T., 1990, in Wheeler, J. C., Piran, T. and Weinberg, S.
{\it  Supernovae} World Scientific Publications.
\ref
[9] Paczy\'nski, B., 1991, Acta Astronomica, {\bf 41}, 257.
\ref
[10] Piran, T., Narayan, R. and Shemi, A., 1992, in
  {\it  Gamma-Ray Burst, Huntsville, 1991}
  Paciesas W. S. and  Fishman, G. J., eds, (AIP), {\it in press}.
\ref
[11] Narayan, R., Paczy\'nski, B., and Piran, T., 1992, \ApJL,
{\bf 395}, L83.
\ref
[12] Clark, J. P. A, van den Heuvel, E. P. J., and
Sutantyo, W., 1979, {\it Astron. and Astrophys.}, {\bf 72}, 120.
\ref
[13] Narayan, R., Piran, T. and Shemi, A., 1991, \ApJL, {\bf 379},
L17.
\ref
[14] Phinney, E. S., 1991, \ApJL, {\bf 380}, L17
\ref
[15] Piran, T. and Shemi, A.,  1993, \ApJL, {\bf 403}, L67.
\ref
[16] Piran, T., 1992, \ApJL, {\bf 389}, L45.
\ref
[17] Paczy\'nski, B., 1986, \ApJL, {\bf 308}, L43.
\ref
[18] Mao, S. and Paczy\'nski, B. 1992, \ApJL, {\bf 388}, L45.
\ref
[19] Dermer, C. D. 1992,  {\it Phys. Rev. Letters,} {\bf 68}, 1799.
\ref
[20] Schmidt, M. 1992, in {\it  Gamma-Ray Burst, Huntsville, 1991},
Paciesas W. S. and  Fishman, G. J., eds, AIP press.
\ref
[21] Krolak, A., 1989, in {\it Gravitational Wave Data Analysis},
 B.F. Schutz, ed., (Kluwer: Dordrecht), 59.
\ref
[22] Peters, P.C., \& Mathews, J., 1963, {\it Phys.Rev.}, {\bf 131}, 435.
\ref
[23] Kochanek, C.S., 1992, {\it Ap.J.}, {\bf 398}, 234.

\bigskip

\noindent{\bf Figure 1:} The figure shows the integral event rate of binary
 coalescences with characteristic strains greater than $h_c$.
 The three models are
 a fit to the GRB rates with a constant comoving rate and $\alpha=1$ spectral
 index (solid line), a fit to the GRB rates with a comoving rate proportional
 to $r(z) \propto t(z)^2$ (dashed line), and a constant event rate of
 $10^{-7} h^{-3}$ Mpc$^{-3}$ yr$^{-1}$ (solid with points).  The strain
 scales as $h_0 \mu_{NS}^{1/2} M_{NS}^{1/3} f_{c2}^{-1/6}$.  The estimated
 LIGO sensitivity $h_{3/yr}$ is shown by the vertical solid line, and the
 sensitivity improvement from using coincidences with GRBs is shown by the
 vertical dashed line.
 strains $h_c$ is shown as a function
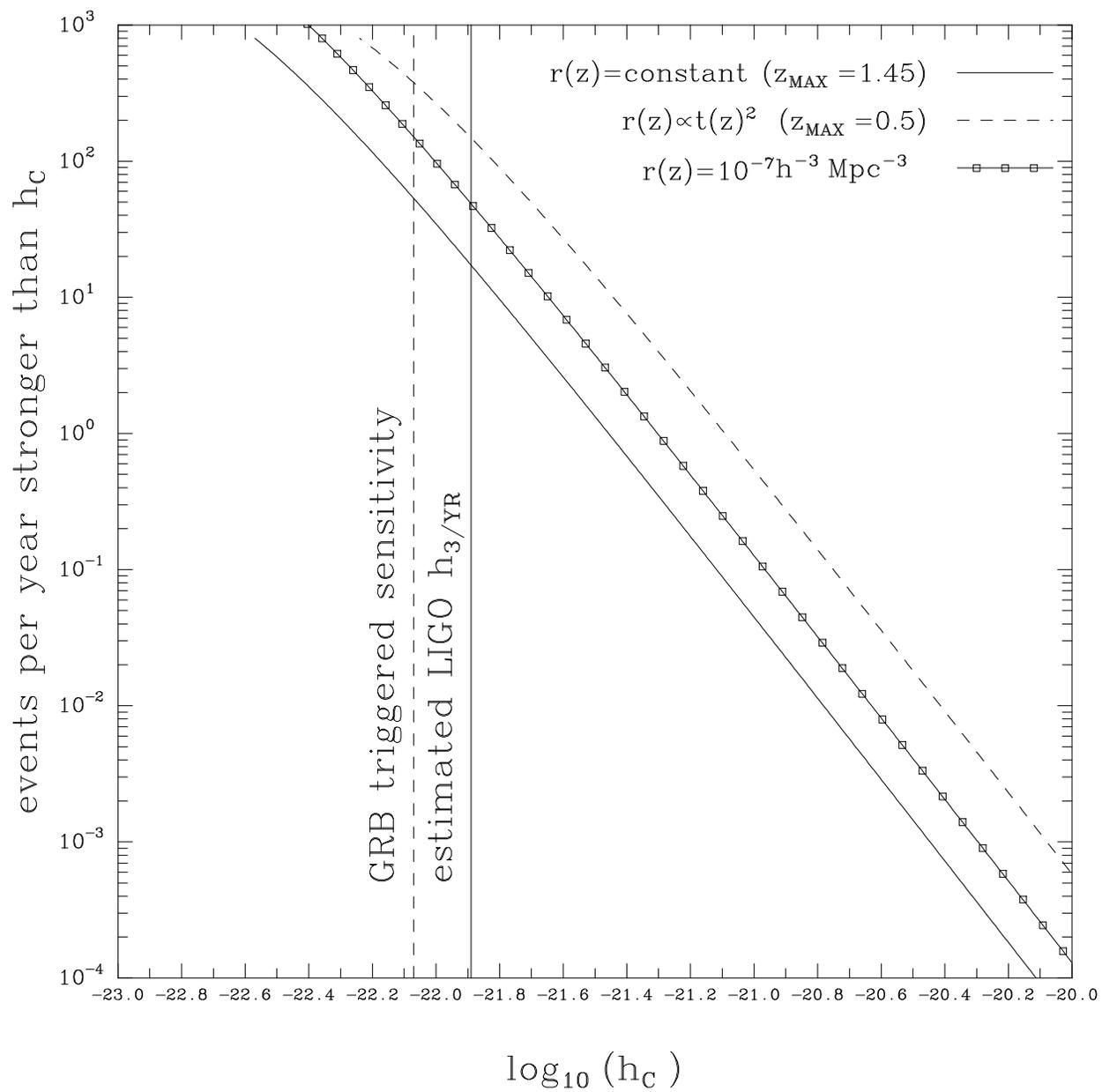
\bye